\documentclass[12pt,preprint]{aastex}

\newcommand{\lsim}{\lower.4ex\hbox{$\;\buildrel <\over{\scriptstyle\sim}\;$}}
\newcommand{\gsim}{\lower.4ex\hbox{$\;\buildrel >\over{\scriptstyle\sim}\;$}}

\begin{document}

\title{Estimating stellar rotation from starspot detection during planetary transits}

\author{Adriana Silva-Valio\altaffilmark{1}}

\altaffiltext{1}{CRAAM, Universidade Presbiteriana Mackenzie, Rua da Consola\c c\~ao, 896, S\~ao Paulo, SP 01302-907, Brazil {\it (asilva@craam.mackenzie.br)}}

\begin{abstract}

 A new method for determining the stellar rotation period is proposed here, based on the detection of starspots during transits of an extra-solar planet orbiting its host star. As the planet eclipses the star, it may pass in front of a starspot which will then make itself known through small flux  variations in the transit light curve. If we are lucky enough to catch the same spot on two consecutive transits, it is possible to estimate the stellar rotational period. This method is successfully tested on transit simulations on the Sun yielding the correct value for the solar period. By detecting two starspots on more than one transit of HD 209458 observed by the Hubble Space Telescope, it was possible to estimate a period of either 9.9 or 11.4 days for the star, depending on which spot is responsible for the signature in the light curve a few transits later. Comparison with period estimates of HD209458 reported in the literature indicates that 11.4 days is the most likely stellar rotation period.

\end{abstract}

\section{Introduction}

Four centuries have elapsed since the first determination of the rotational period of a star, namely the Sun, by observation of the apparent movement of starspots on its surface. For star other than the Sun, the determination of the rotational period is made basically by two methods: either from the rotational broadening of spectral lines or by the periodic modulation of the stellar flux due to the dark and bright features on the stellar surface which rotate with it.
The latter method has the advantage of determining directly the period of a star, without the sin$i$ spectroscopic uncertainty. Moreover, it can also be applied to stars with long rotational period, which cannot be determined from Doppler broadening of spectral lines.

Here I would like to propose a new way of estimating the stellar rotation by using planetary transits. If by any chance, during a transit, the planet passes in front of a starspot, as in Silva (2003), and in a consecutive transit the configuration is such that this same spot is again occulted, then it is possible to estimate the stellar rotation as it was done for the Sun four centuries ago.
 
Previous authors have used a similar method to estimate, among other things, the rotational period of the components of eclipsing binaries. Eaton and Hall (1979) explained successfully the light curve variations of RS CVn type stars as due to starspots. Besides the rotation period and its variation, durations of activity cycles and surface area coverage of spots have also been obtained for the prototypes of active binaries RS CVn (Rodono et al., 1995) and RT Lac (Lanza et al. 2002). 
These are very active stars, nevertheless it will be possible to study moderately active stars such as our Sun with space missions such as CoRoT, MOST, and Kepler. In anticipation to these satellites, 
the rotational modulation of the Sun as a star has been modeled using the VIRGO/SoHO data (Lanza et al. 2003, 2004).

The method described here, however, has a much better spatial resolution than that of eclipsing binaries, because of the smaller planet diameter in comparison to that of the stars. Moreover, planetary transits such as the ones used here will be quite commonly detected by the CoRoT mission already in operation.  
Section \ref{transit} presents the method, which is tested on the Sun with a fiducially transiting planet (Section \ref{solar}) and then applied to HD209458 (Section \ref{hd}). The last section discusses the results of this method and presents the conclusions.

\section{Planetary transit across a starspot} \label{transit}

Similar to Silva (2003), the star is simulated by either a white-light image of the Sun (linear limb darkening) or a model image constructed with the appropriate limb darkening (quadratic in the case of HD 209458), whereas the planet is represented by an opaque disk with a radius of $R_p$, in units of the stellar radius, $R_s$. The planetary orbit is described by its semi-major axis, $a$ (in units of $R_s$) and inclination angle, $i$. To simplify the model, the orbit is assumed to be circular. The light curve is obtained from transit simulations by centering the planet at its given position in the orbit every two minutes and calculating the integrated flux by summing over all the pixels in the image. A search in parameter space of $R_p$, $a$, and $i$ is performed by comparing the simulated light curve with the observed one and the best fit is chosen as the one that minimizes $\chi^2$ (Silva \& Cruz 2006). When the orbital parameters and planet radius are known, however, this is not needed. Assuming that the planetary orbit is in the same plane of the stellar equator, the latitude of the transit chord on the stellar disk is given by $lat = \sin^{-1} [(a/R_s) \cos{i}]$.
The latitude can be arbitrarily chosen to be South (negative) or North (positive).

For the modeling of the spot, three extra parameters are needed: (i) intensity, as a fraction of the stellar (maximum) intensity at disk center; (ii) size, as a fraction of planet radius, $R_p$; and (iii) position, that is, latitude and longitude. The transit latitude is fixed by the inclination angle of the planetary orbit. Only small displacements from the transit latitude are allowed for the spot latitude, measured in $R_p$, otherwise the spot will not be occulted by the planet. Longitude zero is defined as the projection of the planet onto the star during mid transit.

\section{Solar rotation from planetary transits}\label{solar}

\begin{figure}
\plotone{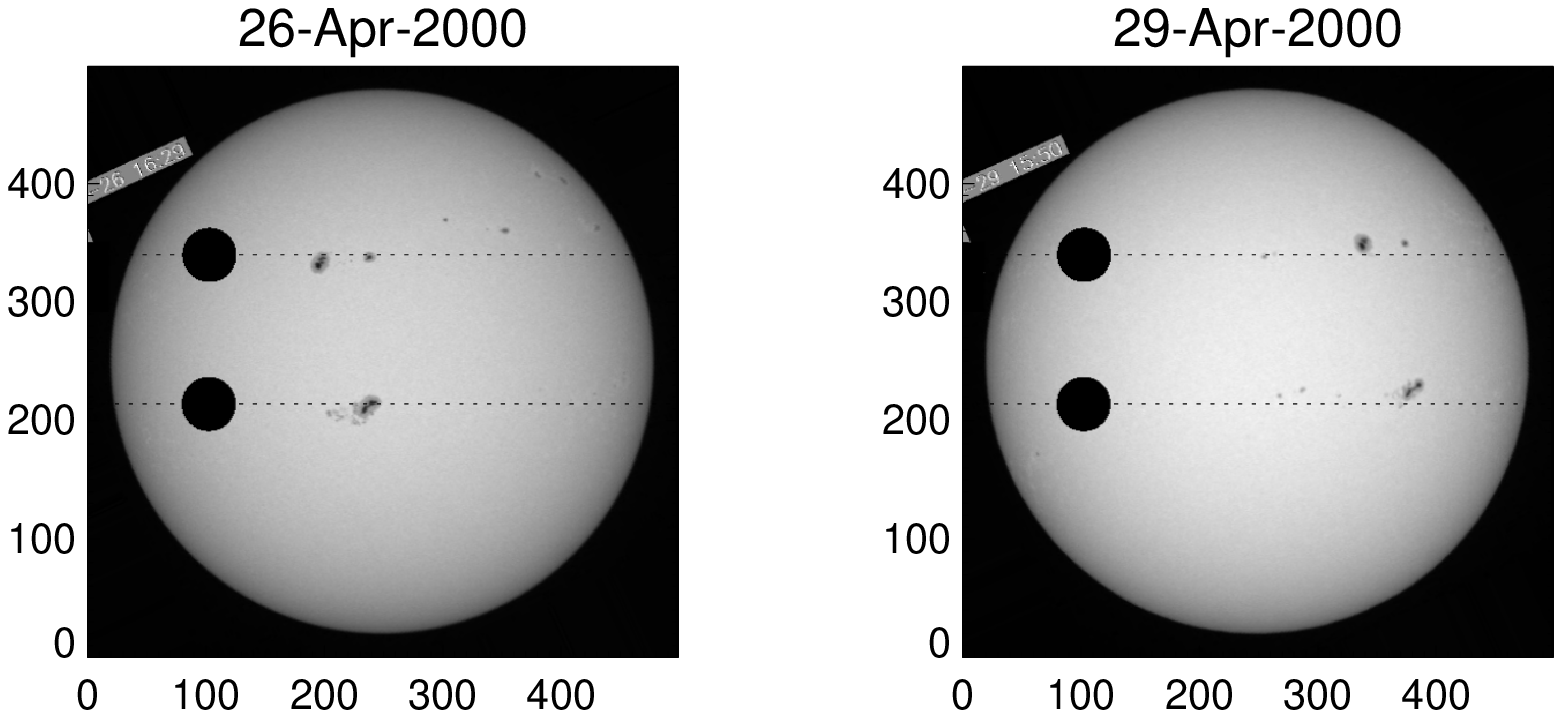}
\figcaption{BBSO images of the Sun on April, 26th (left) and 29th (right), 2000. The simulated transit of a planet at $-9^\circ$ and 23$^\circ$ latitudes are shown as dotted lines.  
\label{suns}}
\end{figure}

As an example of the usefulness of this method, it is tested in the solar case. For this, two images of the Sun are used, taken 3 days apart, on the 26th and 29th of April, 2000 (Figure~\ref{suns}) by the Big Bear Solar Observatory (BBSO). The images have been rotated such that solar North is up. In this simulation, a planet the size of Jupiter ($R_p/R_s = 0.1$) is made to follow an orbit with a semi-major axis of 15 $R_s$. The inclination angle, $i$, is taken such that the planet passes in front of a group of sunspots on both days. Two simulations were performed so that both active regions on the Southern and Northern hemispheres of the Sun are occulted by the planetary transit with inclination angles of -89.4$^\circ$ and $88.5^\circ$, yielding latitude crossings of $-9.0^\circ$ and $23.1^\circ$, respectively, depicted by the dotted lines in the images. 

\begin{figure}
\plotone{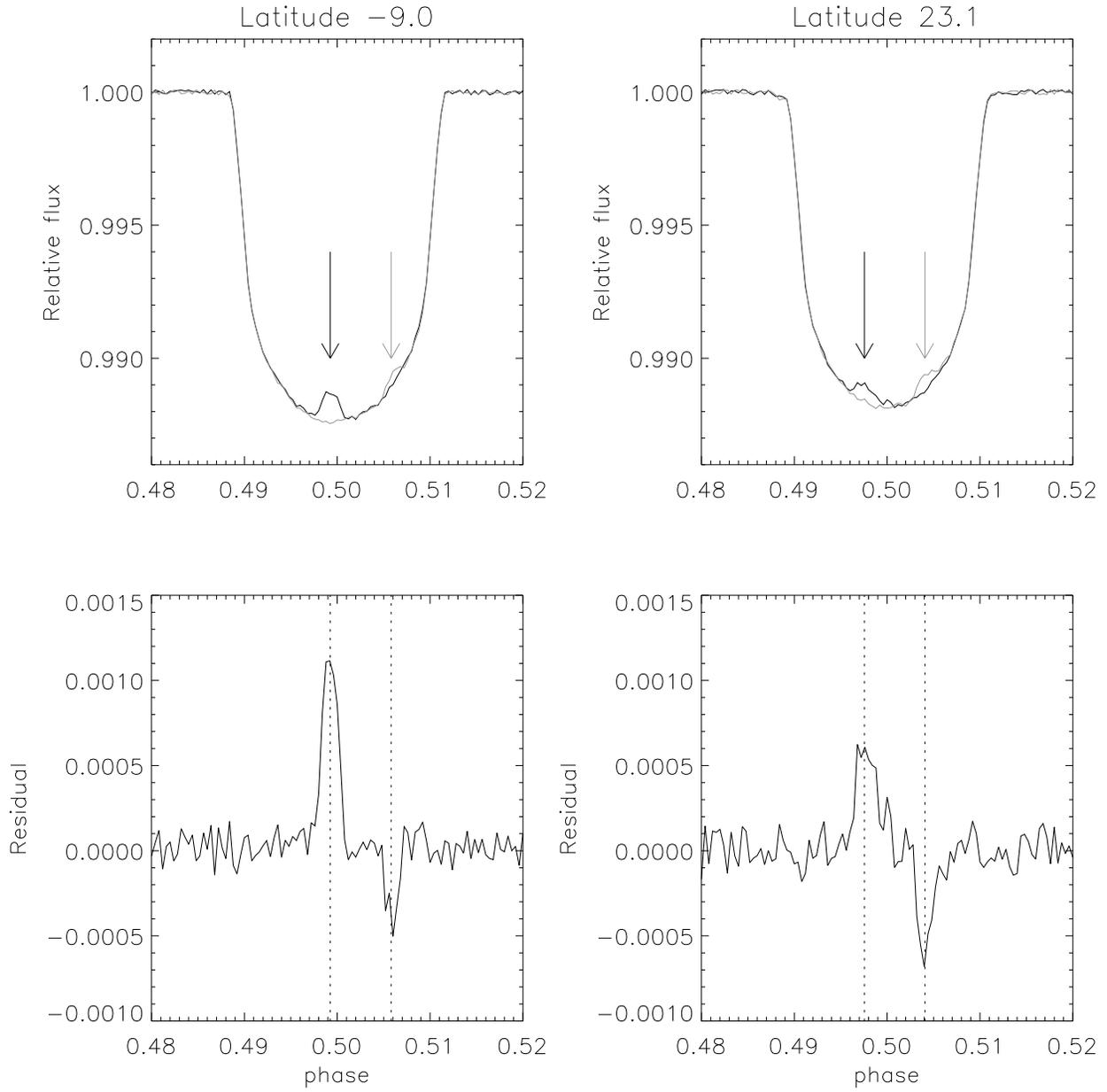}
\figcaption{Light curves of the planetary transit on April 26th (black) and 29th (gray) are shown in the top panels, whereas the bottom panels display the subtraction of the two light curves. The arrows indicate the flux variations caused by the sunspots. The panels on the left are for $-9^\circ$ latitude South and the ones on the right are for 
23$^\circ$.1 latitude North.
\label{sunn}}
\end{figure}

The light curves on each day were calculated, as mentioned in Section~\ref{transit}, by determining the position of the planet every two minutes and summing over all the pixels in the image. To simulate the Hubble Space Telescope observations of a star like HD 209458, a random noise of $10^{-4}$ relative amplitude was added to the model light curve.
The resulting light curves are shown in the top panels of Figure~\ref{sunn} for both days. Because the spots are closer to the limb on April 29th than on the 26th, their signature is less evident on the light curve of the 29th. However, by subtracting one light curve from the other, the signal becomes much more evident, as can be seen in the bottom panels of Figure~\ref{sunn} for both simulations. Note that the effect of the Southern sunspot group on April 26th is about 0.1\% of the total intensity, whereas this value is less than 0.05\% on the 29th. For the Northern sunspot, the values are comparable, being 0.05\% on both days.

Now the rotation period of the Sun at each latitude can easily be calculated from the apparent shift in longitude of the sunspot group after three days. Taking into consideration the surface curvature of the Sun, the stellar period can be determined by:

\begin{equation}
P_s = 2 \pi {\Delta t \over \theta_1 - \theta_2}\\
\label{eq:pstar}
\end{equation}

\begin{equation}
\theta_{i} = \sin^{-1} \left\{  \left[ \frac{a}{R_{\rm s} \cos(lat)} \right]      \sin [2 \pi ( f_{i} -0.5)] \right\} \quad\quad\quad  i=1,2 
\label{eq:long}
\end{equation}

\noindent where $lat$ is the latitude of the transit (-9$^\circ$ and $23.1^\circ$ for these simulations) and $\theta_i$ is the longitude of the spot on the two days considered. These equations assume that orbital plane of the planet is the same as the equatorial plane of the star.
The spot longitude, $\theta_i$, is calculated from the phase, $f_i$, of the positive and negative peaks of the subtracted light curve on the bottom panels of Figure~\ref{sunn} according to Eq.(\ref{eq:long}), and the rotational period of the star, $P_s$, is then estimated from Eq.(\ref{eq:pstar}).

Because of the added noise, however, the exact phase, $f_i$, of the peaks of the subtracted light curves varies a little, thus resulting in slightly different periods. Therefore, the solar rotational period at both latitudes is  obtained by taking the average of a thousand simulations.
The ``stellar" rotation period inferred from these simulations is 27.3 $\pm$ 0.6 days for the Southern spot and 27.7 $\pm$ 0.7 days for the Northern one. These values agree quite well with the values obtained directly from measuring the longitude of the spots on the solar images, which yields 27.23 and 27.65 days, respectively, for both sunspot groups. 
As can be seen from these values, there is a small difference in the rotational period obtained from the spots at different latitudes, because of the known solar differential rotation. Thus it might be possible to measure differential rotation in other stars if the planet is large enough to cover simultaneously spot groups at different latitudes during the same transit.

\section{Rotation period of HD 209458}\label{hd}

\begin{figure}
\plotone{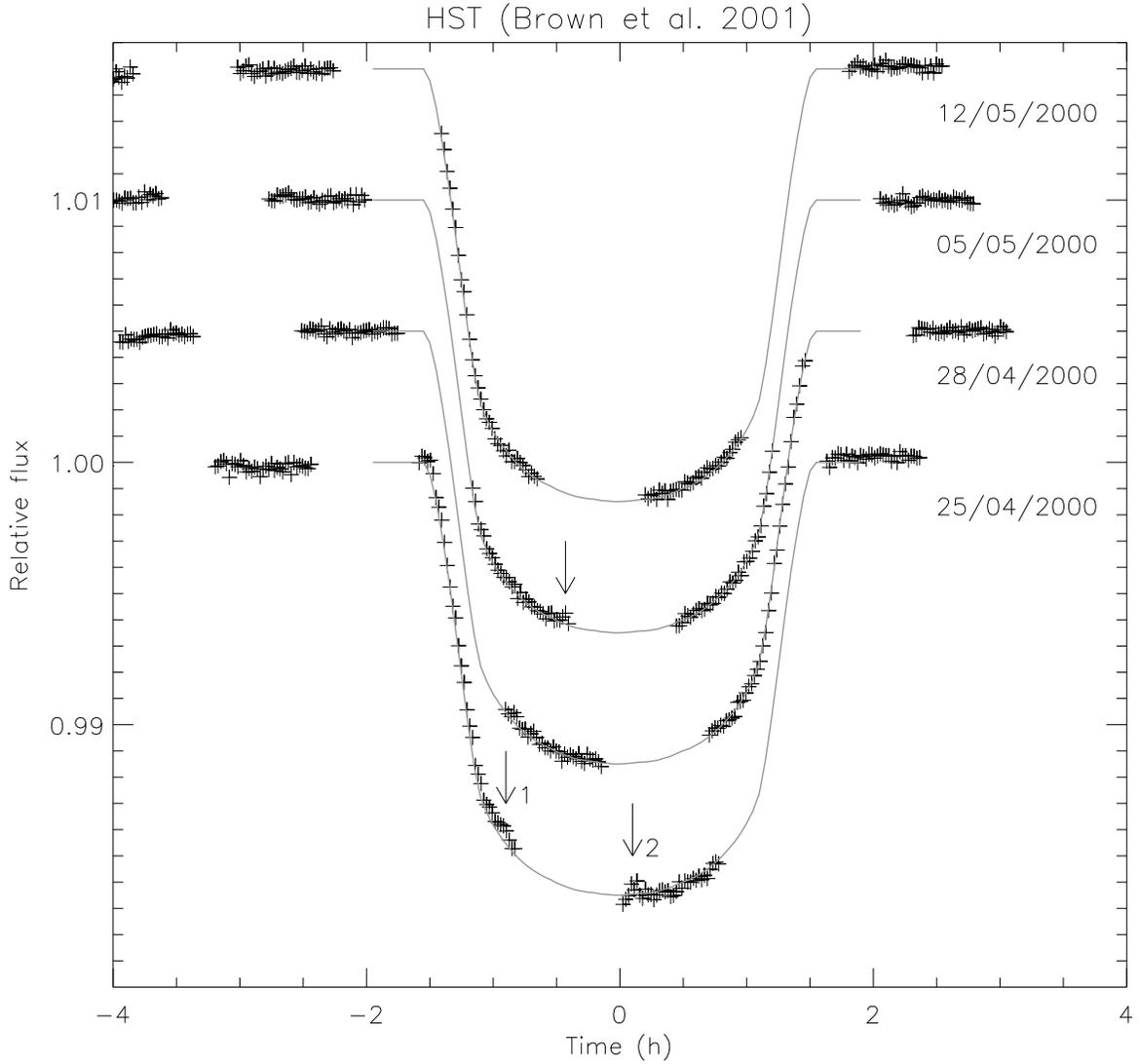}
\figcaption{HST observations of four transits of HD 209458b on April and May 2000. Deviations from the light curves, interpreted as due to the presence of starspots, are indicated by arrows on two of the transits. 
\label{hdlc}}
\end{figure}

Next I apply this same method to the observations of HD 209458 by the Hubble Space Telescope (HST, Brown et al. 2001). HST observed four transits of HD 209458b on April and May of 2000 (see Figure~\ref{hdlc}). From the first transit on April 24th until the last one on May 12th, five orbital periods have elapsed, the orbital period of HD 209458b being 3.52475 days (Brown et al. 2001). The April 25th light curve presents two ``bumps" interpreted as due to two spots, indicated by the arrows. The first one at about $t = -1$ h before transit center is due to spot 1, whereas the second flux variation near transit center ($t = 0$ h) is attributed to the presence of spot 2 located near disk center. The latter spot is the one modeled in Silva (2003).  The goal here is to determine the stellar rotation period of HD 209458 such that one of these spots is exactly at the position required to cause the ``bump" on the May 5th transit (indicated by the arrow), three planetary orbits later, that is, 10.5 days afterwards.

\begin{figure}
\epsscale{.9}
\plotone{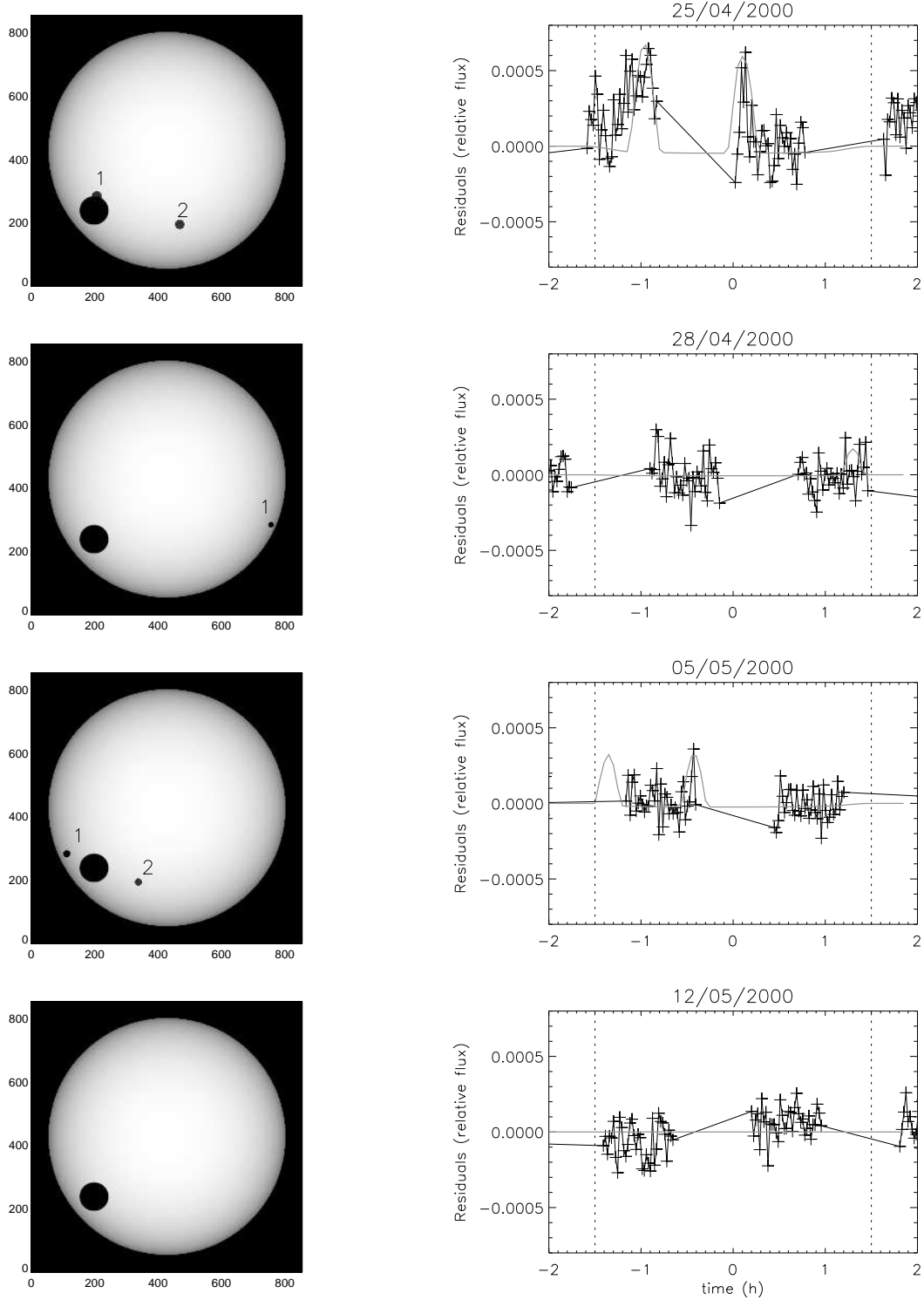}
\figcaption{{\it Left}: Model of quadratic limb darkened star with spots. {\it Right}: Residuals of the four HST transits (crosses), whereas the gray solid line represents the residuals from the model considering a stellar rotation period of 11.4 days.
\label{per11}}
\end{figure}

Unfortunately, due to gaps in the HST data, it is not possible to subtract one light curve from the other as was done in the solar case. The light curve of the transit without spots was calculated using the known parameters for HD 209458b, that is, a  planet with radius $R_p = 0.1366 R_s$, inclination angle of 86.68$^\circ$, and orbital radius of 0.04656 A.U. which is equivalent to 8.78 $R_s$. The light curve obtained by applying this model to a star with quadratic limb darkening [$I(\mu)/I(1) = 1 - w_1(1-\mu) - w_2(1-\mu)^2$, where $\mu=\cos\theta$, $w_1 = 0.2925$, and $w_2 = 0.3475$, Brown et al. 2001] and without any spots is shown as a solid gray line in Figure~\ref{hdlc}.
This spotless light curve is then subtracted from each observed transit. Figure~\ref{per11} shows the result of this subtraction as crosses on the right column plots. 

Initially, the spots are located at latitudes and longitudes (-22.5$^\circ$, -40$^\circ$) for spot 1 and  (-38.4$^\circ$, 8$^\circ$) for spot 2. Their latitudes were shifted from the transit line by +1 or -1 $R_p$ to better distinguish between them. During the first transit, the spots were modeled with radii of 0.33 and 0.36 $R_p$, respectively, and with 0.3 of the stellar central intensity. These spots can be seen in the left top most panel of Figure~\ref{per11}. A simulated light curve is produced by making the planet transit in front of these spots, and then the  spotless light curve is subtracted from it. The result is shown as the gray curve on the top right panel of the figure.

To obtain the period of the star, the position of the spots after each transit are calculated by considering the star to rotate with a certain period, $P_s$, for the appropriate time interval between transits (multiples of 3.52475 days). The light curve obtained from this model is then compared with the observations. By varying $P_s$ in steps of 0.05 days, the best value was chosen as the one where the ``bump" at $t = -0.43$ h, coincided with the peak of relative intensity of 0.0004 on the May 5th transit. For comparison, the noise of these data, measured by the standard deviation of the residuals of the $t>0$ points, is 0.00010. 

For spot 2, the best match between the times of the modeled and observed peaks was obtained for a stellar period of 11.4 days. The residuals of the light curves for the four transits are shown on the right panels of Figure~\ref{per11}, the data as crosses and the model as a solid gray line. For a 11.4 days stellar period, spot 2 was on the other side of the star during the transit of April 28th, and both spots were behind the stellar limb during the last transit on May 12th. Moreover, in order to fit the peak intensity on May 5th, which is smaller than that on April 25th, it was necessary to decrease the size of spot 2 from 0.33 to 0.25 $R_p$. This may be an indication of spot evolution, decaying in size after approximately 10 days.  

\begin{figure}
\epsscale{.9}
\plotone{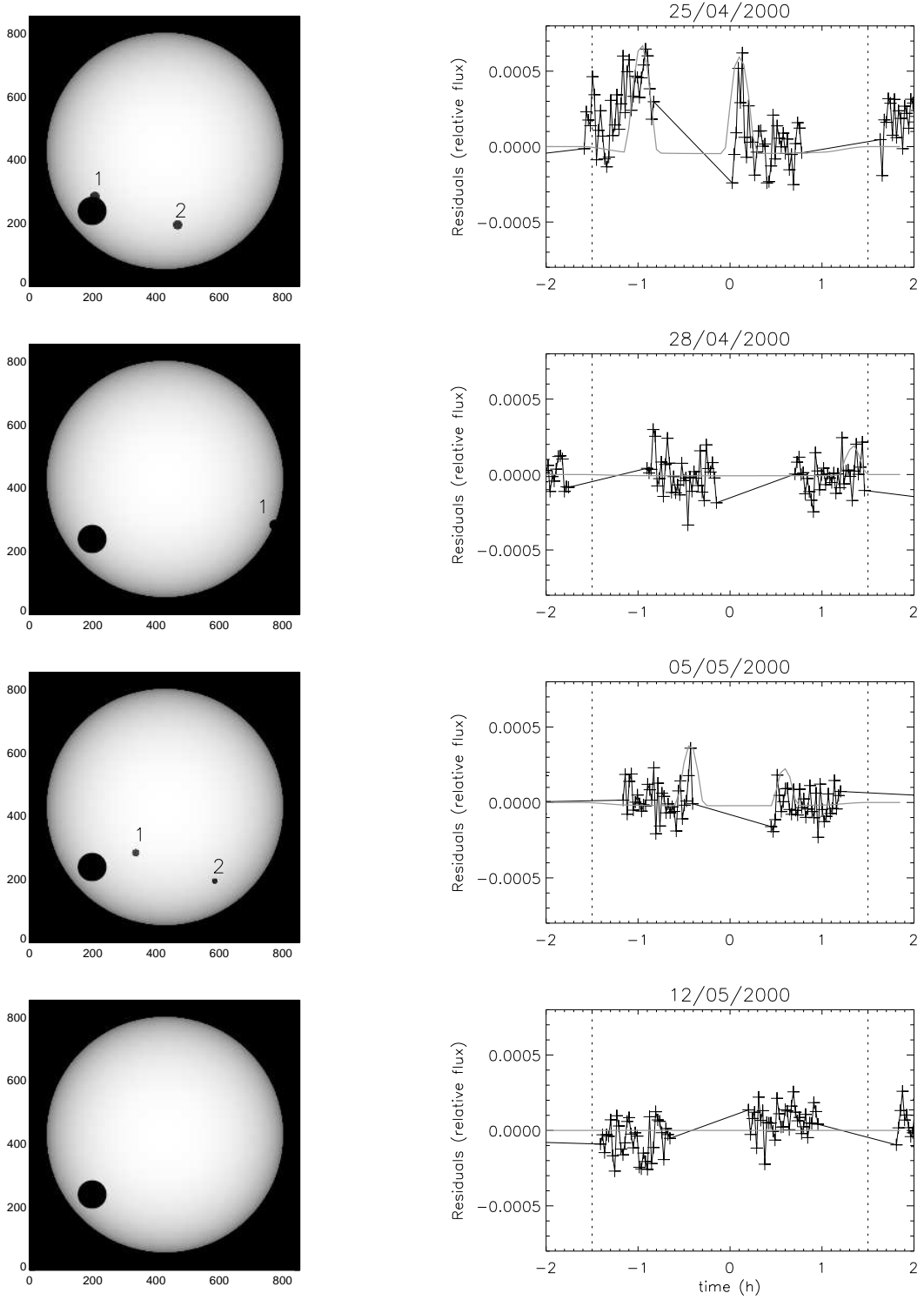}
\figcaption{Same as Figure~\ref{per11} for a stellar rotation period of 9.9 days.
\label{per9}}
\end{figure}

What if the signature in the May 5th transit light curve is caused by spot 1 instead of spot 2, which means that the star rotates faster than 11.4 days? The same procedure was repeated, by varying $P_s$ such that the ``bump" caused by spot 1 turned up at $t = -0.43$ h on May 5th. This was successfully done for a stellar rotational period of 9.9 days, and the results are shown in Figure~\ref{per9}. Also in this case, the spot is seen to decrease in size from 0.36 to 0.27 $R_p$. 

Unfortunately, due to the gaps in the data it is not possible to distinguish between the two cases: 9.9 or 11.4 days for the rotational period of HD 209458. If there were complete data coverage, it might be possible to distinguish between the two period values using the data just after ingress on the May 5th transit, since the model for the 11.4 days predict a spot signature there whereas the light curve for the 9.9 days period does not.

\section{Discussion and Conclusions}

This work proposes to estimate the rotational period of a star by following the apparent shift in longitude position of its surface spots, similar to what was done by Galileo and his contemporaries four centuries ago for the Sun. Tracking of the spots is done by identifying ``bumps" in the light curves of successive planetary transits.

This method was successfully tested for the Sun yielding the correct value for the solar period from simulated transits taken only three days apart even though the solar period is about 27 days. Moreover, by modeling of different sunspot groups it was possible to verify the solar differential rotation.
Supposing that the small variations detected in the light curve during planetary transits is due to occultation of starspots, the model was also applied to HD 209458 using HST observations obtained by Brown et al. (2001) in April and May of 2000. The star was modeled as having two spots during the four transits. The ``bumps" detected on two transits separated by three orbital periods yield a stellar rotation period of 9.9 or 11.4 days depending on which of the spots detected on April 25th is considered to cause the intensity variation on the May 5th transit light curve.

Several observations of $v \sin i$ of HD 209458 are found in the literature. Assuming that the inclination angle is 86.68$^\circ$ and the stellar radius 1.148 R$_\odot$, these observations yield stellar periods of 14.4 $\pm$ 2.1 days (Mazeh et al. 2000) and 15 $\pm$ 6 days (Queloz et al. 2000). More recently, shorter periods have been found, 12.3 $\pm$ 0.5 days (Winn et al. 2005) and 12 days (Fisher \& Valenti 2005). 

The periods obtained here, 9.9 and 11.4 days, are a little shorter than those found in the literature. It seems that the 11.4 days period is the true one for HD 209458, even though it is still a little shorter than the periods listed in the literature.
However, HD 209458 probably presents differential rotation, similar to the Sun. In this case, the period obtained by other authors, which was based on line broadening observations, is actually an average of the periods of the whole stellar disk, whereas the period determined here represents the rotational velocity at that specific latitude which is close to the equator. 
As the planet size spans about 20$^o$ in stellar latitude, we are probing latitudes from -22$^o$ to -38$^o$. Since the work of Winn et al. (2005) was based on the observation of the Rossiter-McLaughlin effect during transits, therefore obtained at the same latitudes we sample, this is the result which should better agree with the one presented here. 

Thus far, over 50 transiting planets have been detected (The Extrasolar Planets Encyclopaedia - http://exoplanet.eu) and many more are expected especially in the next months thanks to observations by the CoRoT satellite. The method proposed here can easily be applied to the planetary transits of newly discovered planets and checked against periodic modulation of the stellar flux outside transits themselves.

\acknowledgments

I thank the referee Antonio Lanza for useful comments which much improved the paper. This research was partly supported by the Brazilian agency FAPESP (grant number 06/50654-3).

\clearpage

\end{document}